# Superconductive Excitations and the Infrared Vibronic Spectra of BSCCO

J. C. Phillips

Dept. of Physics and Astronomy,

Rutgers University, Piscataway, N. J., 08854-8019

ABSTRACT

The oxygen dopant concentration dependence of the "quasiparticle" feature at 600-750 cm$^{-1}$ of the infrared spectrum of $Bi_2Sr_2CaCu_2O_{8+\delta}$ (BSCCO), lying near and above the top of the host phonon spectrum at 600 cm$^{-1}$, is strongly correlated with superconductivity. Using parameter-free topological methods, theory assigns this feature to split apical oxygen interstitials. It explains both the qualitative similarities and the quantitative differences between "quasiparticle" features identified in infrared and photoemission data, as well as identifying new features in the infrared spectra.

PACS indices: 74.72.-h  63.20.Kr  74.25.Kc  78.30.-j

## 1. Introduction

High temperature superconductivity (HTSC) is perhaps the most surprising and *a priori* unlikely phenomenon discovered in the last half century. It occurs not in metals, but only in moderately doped creramic cuprates with complex crystal structures and a high level of multiphase nanoscale disorder; this suggests that complexity is the major factor responsible for HTSC. Biological systems are much more complex and exhibit even richer phenomena, often described in terms of self-organization. In complex inorganic systems self-organization can occur in a variational context, for instance, in network glasses (such as window glass) the network relaxes to fill space optimally. Topological theories are particularly well suited [1] to dealing with such complexities, and they typically can incorporate variational properties (such as space-filling) that are not readily described by analytic or algebraic methods involving toy Hamiltonians. Topological analysis of global and local elasticity in undoped host oxide networks has recently explained why HTSC occurs in the cuprates, while colossal magnetoresistance, CMR, occurs in the manganites [2]. In contrast to mean field or Monte Carlo methods involving toy Hamiltonians to treat multiphase complexity, the analysis involves *no adjustable parameters*. Moreover, the general feature of the phase diagrams of both HTSC and CMR, that these phenomena involve intermediate phases that share many anomalous properties (for instance, the normal-state temperature dependencies of the normal and Hall resistivities), are readily explained, again *without adjustable parameters*, by recognizing the distinctive topological properties that create such phases [3]. These properties lie outside the range of mean field theory. They can be generated either by Monte Carlo simulations of toy models with many adjustable parameters [4,5,6], or topologically *without adjustable parameters* by utilizing analogies to network glasses supported by general criteria (such as space-filling) [1-3].

The conventional mechanism responsible for superconductivity is electron-phonon interactions, but in the cuprates there is an abundance of lavishly parameterized models based on other mechanisms. The complexity of the cuprates has made it difficult to prove, using the same methods as were successful for metals (isotope effect, for example), that electron-phonon interactions are the correct mechanism. A large part of the problem here is that metals can be treated quite well by mean-field methods, but HTSC in the cuprates occurs only upon doping, and the dopants must surely form self-organized networks not describable by mean field theory. Strong evidence for such self-organized internal structures was obtained quite early in neutron spectra, which exhibited vibronic anomalies not describable by any mean field model [7]. These anomalies, which occur in the middle of the vibrational spectrum, near 50 meV (400 - 450 cm$^{-1}$), have been the subject of continuing studies, by not only neutron spectroscopy, but also infrared [8] and Raman. The results are described quite well topologically (*without adjustable parameters*) [9,10].

The topological description of dopant networks couples them to internal strain fields, and identifies them with percolative elastic backbones. In contrast to conventional percolation theory, usually based on lattices and random site occupations, the space-filling dopant networks in HTSC are regarded as glassy and off-lattice, and self-organized to maximize dielectric screening energies, as well as minimize internal strain energies. Many features of HTSC that appear from the perspective of mean field theory to be unlikely or accidental, especially composition dependencies in the intermediate phase, are in fact expected in the context of variational self organization, and can be taken as evidence for its presence.

## 2. Host and Doped Phonon Spectra

It is important to appreciate that dopants in cuprates produce very large changes in phonon spectra that cannot be described by conventional spring constant models based on mean-field (the analogue of rigid-band electronic) models. These changes vary with probe (neutron [7] or infrared [8]) and are best understood semiquantitatively using topological methods [9,10]. They are largest near the middle of the spectrum (roughly speaking, between the cuprate acoustic and optic bands), and have already been analyzed there theoretically [9,10], with respect to $YBa_2Cu_3O_{6+x}$ and $La_{2-x}Sr_xCuO_4$, where large single crystals samples are available. The general conclusion is that the midband anomalies consist primarily in localized longitudinal optic modes near the zone boundary that have split off below the bulk LO modes because of an attractive interaction with the dopants.

Mixing of longitudinal and transverse modes in the midband region produces several extra bands with complex compositional trends. Recently analysis of infrared data for $Bi_2Sr_2CaCu_2O_{8+\delta}$ (BSCCO) near and above the top of the host phonon spectrum (optic modes) at 600-750 $cm^{-1}$ has shown a single strong band with a very simple compositional (six single crystal samples) trend with $\delta$ that parallels $T_c$ ($\delta$) [11], so that these trends must be associated directly with the $O_\delta$ dopants. At the top of the phonon band the level of degeneracies is reduced, and this may contribute to the similarities observed between the δ dependencies of these features and the superconductive $T_c$. Also there is a less obvious reason for the correlation, namely certain special structural features of BSCCO, that will become apparent in the Section 4.

There has been considerable confusion concerning the phase diagrams of the cuprates and the effects of multiphase complexity on physical observables; the infrared data discussed here have already been described as "strongly support[ing] the magnetic-resonance interpretation of the self-energy peak" [12]. Of course, it is not easy to see how infrared data can be directly connected to a magnetic resonance (oscillator strength smaller by a factor $(v/c)^2 \sim 10^6$) studied with spin-polarized neutrons. In fact there are several correlations between superconductivity and magnetic properties; however, these arise not from any intrinsic relation between magnetism and superconductivity, but are merely the consequence of large filling factors for interfaces between the two kinds of nanodomains [13]. Our results do not concern such confusion [12], but are directed towards showing that the data can be interpreted using no exotic constructs, only conventional electron-phonon interactions, contrary to the authors' assessment [11] that their "results rule out both the magnetic resonance peak [12] and phonons as the principal cause of high-$T_c$ superconductivity".

In what follows it is assumed that the reader is familiar with the general topological arguments for the filamentary threading nature of the intermediate phase, as discussed in 1990 and more recently in some seventeen of the author's papers on cuprates in the last seven years, obtainable by searching the Web of Science on "phillips jc" + "filamentary". Briefly, currents are carried by nonoverlapping filaments that thread between metallic (in BSCCO, bismuthate and cuprate) planes, through defects, which may often be centered on apical oxygen sites. The defects generate localized states at the Fermi energy; these states resonantly mix with some of the states in the metallic planes to generate filamentary paths. The electron-phonon interactions are supposed to be weak in the stiff cuprate planes, larger in the softer secondary metallic planes, and possibly largest at the interplanar defect bridges. The metallic planes are only locally metallic; because of interplanar misfit, they contain nanodomains that are separated by insulating walls. Thus the filamentary currents must percolate along paths that thread between planes and around intraplanar obstacles (domain walls). This requires a minimum concentration of dopants to reach the percolation threshold; it also implies a maximum concentration, where the filaments overlap and merge to form a Fermi liquid, where mean field theory is appropriate. These two concentrations topologically define the boundaries of the intermediate phase, and compositional fluctuations between these two limits give rise to the broad, rather flat parabola for $T_c(\delta)$ that is used to define $\delta$ and is reflected in the infrared spectrum.

In the cuprates the infrared spectrum shows large currents in the ab planes, upon which the phonon spectrum is superposed. To extract the single-particle energies [11] uses a Drude self-energy model that incorporates lifetime ($\tau$) broadening effects. [Strictly speaking, it would have been more accurate to use a fractal Drude model [14] with $\eta = 3/4$ instead of $\eta = 1$ [15], but this refinement would probably not have changed the results much.] With decreasing temperature the observed spectra sharpen as the scattering rates $\tau^{-1}$ decrease, leading to an apparent enhancement of the maximum self energy $\omega_1$ (presumably the low temperature limits measure the intrinsic self energies most accurately) to around 750 cm$^{-1}$ at optimal doping. This enhancement includes a contribution from a broad background that exists above the cutoff in the phonon spectrum, as measured by neutron scattering. [11] then argues that although "the broad background can be seen in the spectra of all high-$T_c$ superconductors and seems to represent a universal property of the copper oxide plane", "it cannot be due to phonons, because its spectral weight extends beyond the cut-off frequency of the phonon spectrum" (around $\omega_c = 600$ cm$^{-1}$). The latter, however, has been determined using a probe (neutrons) that does not excite currents. The important point here is that the maximum value of $\omega_1$ (achieved at optimal doping, maximized $T_c$) is only 25% above the host phonon cutoff $\omega_c$ determined by neutron scattering. $\omega_1$ is weighted more heavily in the neighborhood of the dopants in the filamentary model because filamentary currents pass resonantly through dopant bridges. The same point was made earlier in more conventional polaron language [16,17]. The extra absorption just above $\omega_c$ is not only explained, but is actually expected, providing that there is a dopant-centered phonon (or polaron) resonance that has split off from the top of the host phonon band. Our only task now, but it is not a small one, is to determine the nature of the defect, from which it will be easy to estimate its highest resonant frequency.

### 3. YBCO and BSCCO Phonon Spectra by Neutron Scattering

The previous discussions [9,10] of cuprate phonon spectra compared and contrasted infrared and neutron spectra in great detail, with emphasis mainly on YBCO [YBa$_2$(Cu$_{1-y}$Zn$_y$)$_3$O$_{6+x}$], because it is there that the data base is most complete. The changes in the spectra with x and y are very large, and as emphasized in the experimental papers, are simply inexplicable with conventional force-constant models based on ideal lattices. It is plain that the superconductive phases must contain large internal structural changes that can be described most easily and most generally in topological terms, that is, the changes with x are associated with formation of filamentary connectivity by self-organized oxygen dopants, while the even larger changes with y ~ 0.04 are caused by disruption of these filaments by substitution of a few % Zn impurities for Cu in the CuO$_2$ planes. [Note that large changes of structure and T$_c$ with small y are inexplicable using popular models that ascribe HTSC to interactions between carriers in the CuO$_2$ planes only.]

The largest changes in the YBCO neutron spectra occurred between 300 and 400 cm$^{-1}$, and those were discussed in detail [9,10]. However, important changes also occurred near the top of the phonon band. Without O in the secondary Cu plane (x = 0) there are two peaks near 72 and 80 meV associated with apical O and planar CuO$_2$ (Cu-O) longitudinal optical stretches, respectively. The latter peak seems to disappear when the CuO chains are formed (x =1), but it seems more likely that it softened and shifted to 70 meV due to screening by metallic carriers associated with filaments that include CuO chain segments. Thus the filamentary model explains the startling reappearance of a peak near 85 meV with 10% Zn substitution. In classical models the main effect of the Zn should be contained in the mass change, which is small and which should have lowered the phonon energy, not increased it. However, in the filamentary model, the Zn disrupts the topological connectivity of the filaments, and greatly reduces screening by metallic carriers.

The phonon spectra of $Bi_2Sr_2(Ca_{1-x}Y_x)Cu_2O_{8+\delta}$ were studied with neutron scattering on powder samples [18] by varying x, with $\delta$ adjusted for x = 0 to give $T_c \sim 80$ K; the value of $\delta$ for x = 0.6 and 1 ($T_c \sim 0$). Although the changes from x = 0 to x = 1 here were smaller, they still could not be explained as mass effects, as these should have occurred by shifting a peak near 25 meV to lower energies. Instead the main changes observed were enhancement of peaks at 20 and 40 meV and reduction of a 30 meV peak, and broadening of a 72 meV LO peak over a range from 65 to 80 meV, suggestive of Jahn-Teller distortions in the $T_c \sim 0$ samples.

## 4. Dopant Complexes in BSCCO

Our search for the nature of dopants in BSCCO is considerably simplified by the fact that we know they must involve $O_\delta$. These dopants must be *somewhere* in the lattice, but they have never been identified in the ultrahigh resolution STM studies [19]. However, these impurities may be highly mobile and so not observable on the laboratory time scale and in the presence of tunneling probe currents. Dilute secondary dopants can fix the positions of primary dopants in their neighborhood. The STM studies exhibit rich subgap structure in the vicinity of secondary dopants like Ni and Zn, that is not explained by central site perturbing potentials; this subgap structure is centered on Cu or Bi or apical oxygen sites collinear with Cu and Bi along the c axis Therefore [13] suggested that the natural site for $O_\delta$ is a split apical O interstitial; it also noted that split anion interstitials are characteristic of materials with small (first period) anions and large cations that provide a convenient "shadow" under which the split interstitial is stable.

A model of $O_\delta$ dopant complexes as split apical interstitials is thus consistent with the STM data and with crystal chemical trends. It also makes several predictions that can be easily tested against the observed spectra [11]. In a marginally stable elastic network [2], equilibrium conditions require approximate equality of local atomic forces. The highest frequency $\omega_D$ of an O-O pair scales with its reduced mass $\mu_D$ against $\mu_H$, the reduced mass of the host Cu-O LO mode, $\omega_H$. Thus $\omega_D^2 \mu_D = \omega_H^2 \mu_H$, and with $M_{Cu} = 4 M_O$, $\omega_D = 1.26 \omega_H$. The maximum LO neutron peak energy is ~ 75 meV = 600 cm$^{-1}$ in Bi$_2$Sr$_2$CaCu$_2$O$_{8+\delta}$ (BSCCO), so the maximum frequency for an LO defect mode based on O-O pairs is ~ 750 cm$^{-1}$. The maximum value obtained in the extended Drude analysis of the infrared spectra at optimal doping is ~ 750 cm$^{-1}$, so the agreement here is excellent. Moreover, the filamentary enhancement model explains the observed composition dependence perfectly: at the edges of the superconductive phase the filamentary effects are small, and the maximum frequency reverts to that of the neutron spectra, ~ 600 cm$^{-1}$.

Is that all? No, there's more in the LO split apical oxygen interstitial (polaron) model. Such split interstitials are both highly mobile and only marginally stable. Hence one would expect to observe large anharmonic effects, but due to the broadening effects of interactions among the dopant centers, these will be most easily observed in underdoped spectra (Fig. 1e of [11]). There we see a broad but unmistakable harmonic between 1000 cm$^{-1}$ and 2000 cm$^{-1}$, centered near 1500 cm$^{-1}$, and if one is optimistic, there is even a second harmonic above 2000 cm$^{-1}$ centered near 2250 cm$^{-1}$.

## 5. Comparison with ARPES

In Fig.2b of [11] the composition dependencies of the maximum infrared frequency and the phonon kink observed by ARPES [20] are shown to be similar, but they apparently differ by a factor of ~ 2. It seems likely that the ARPES spectra are probably more strongly coupled to the midband filamentary vibronic modes evident in the neutron spectra near 50 meV, as discussed above.

## 6. Conclusions

Here the author has continued his straightforward topological analysis, based on the filamentary concepts in some seventeen of the author's papers on cuprates in the last seven years, obtainable by searching the Web of Science on "phillips jc" + "filamentary". This analysis shows, without utilizing adjustable parameters or invoking mystical concepts such as "strongly correlated electronic glues", that the neutron, ARPES, and infrared spectra are all quantitatively mutually consistent with a model of dopant-mediated strong electron-LO phonon interactions in the cuprates. This parameter-free model represents a microscopic realization of some of the intuitive factors (strong electron-lattice interactions and lattice instabilities in ferroelastic oxides in general and perovskites in particular) that guided Bednorz and Mueller in making their historic discovery [21], a discovery not emulated by theorists with parameterized Coulombic or magnon models.